\documentclass[]{aa}

\usepackage{graphicx}
\usepackage{amssymb,amsmath}
\usepackage[varg]{txfonts}
\usepackage{natbib}
\usepackage{hyperref}
\hypersetup{pdftitle = The title of my PDF, pdfauthor = My name, pdfsubject= The subject, pdfkeywords = keyword1 keyword2 keyword3}
\hypersetup{colorlinks = true, linkcolor = blue, anchorcolor = red, citecolor = blue, filecolor = red, pagecolor = red, urlcolor = blue}

\newcommand{\orcid}[1]{\href{https://orcid.org/#1}{\includegraphics[width=8pt]{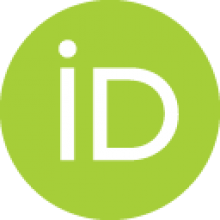}}}

\begin{document}

\title{Orbit classification in a disk galaxy model with a pseudo-Newtonian central black hole}

\titlerunning{Classification of orbits in a disk galaxy with a black hole}
\authorrunning{E.E. Zotos et al.}

\author{Euaggelos E. Zotos\inst{\ref{inst1},\orcid{0000-0002-1565-4467}} \and Fredy L. Dubeibe\inst{\ref{inst2},\orcid{0000-0002-0793-3255}} \and Andr\'{e} F. Steklain\inst{\ref{inst3}} \and Tareq Saeed\inst{\ref{inst4}}}

\institute{
Department of Physics, School of Science, Aristotle University of Thessaloniki, 541 24, Thessaloniki, Greece \\
\email{evzotos@physics.auth.gr}\label{inst1}
\and
Facultad de Ciencias Humanas y de la Educaci\'on, Universidad de los Llanos, Villavicencio, Colombia \\
\email{fldubeibem@unal.edu.co}\label{inst2}
\and
Academic Department of Mathematics, Universidade Tecnol\'{o}gica Federal do Paran\'{a} (UTFPR), \\
Av. Sete de Setembro, 3165, Curitiba, Brazil \\
\email{steklain@utfpr.edu.br}\label{inst3}
\and
Nonlinear Analysis and Applied Mathematics (NAAM)-Research Group, Department of Mathematics, \\
Faculty of Science, King Abdulaziz University, P.O. Box 80203, Jeddah 21589, Saudi Arabia \\
\email{tsalmalki@kau.edu.sa}\label{inst4}
}


\abstract{
We numerically investigate the motion of stars on the meridional plane of an axially symmetric disk galaxy model, containing a central supermassive black hole, represented by the Paczy\'{n}ski-Wiita potential. By using this pseudo-Newtonian potential we can replicate important relativistic properties, such as the existence of the Schwarzschild radius. After classifying extensive samples of initial conditions of trajectories, we manage to distinguish between collisional, ordered, and chaotic motion. Besides, all starting conditions of regular orbits are further classified into families of regular orbits. Our results are presented through modern color-coded basin diagrams on several types of two-dimensional planes. Our analysis reveals that both the mass of the black hole (in direct relation with the Schwarzschild radius) as well as the angular momentum play an important role in the character of orbits of stars. More specifically, the trajectories of low angular momentum stars are highly affected by the mass of the black hole, while high angular momentum stars seem to be unaffected by the central black hole. Comparison with previous related outcomes, using Newtonian potentials for the central region of the galaxy, is also made.
}


\maketitle

\section{Introduction}
\label{intro}

Galactic dynamics is of paramount importance not only because of the evident astronomical interest in classifying and analyzing the nature of galaxies \citep{BT08, MA11,Z12}, but also to investigate the nature of dark matter that constitutes a significant part of these objects \citep{ZCar13}. Exploring the nature of orbits on different models of galaxies can provide a hint to the nature of dark matter present in these structures. Moreover, the phase space structure of a particular model shall depend on its basic components, i.e., disk, halo, and the nucleus or central bulge \citep{Z12,ZC13,Z14}, such that its phase space is as unique as a fingerprint. In general terms, orbits can be classified according to their dynamical behavior into regular and chaotic \citep{C96,CP03,Z12,ZC13,Z14}, while regular orbits can be subclassified into different families depending on its resonances \citep{MM75,M79,G87,LS92}, which can be analyzed using different technics \citep{BS82,BS84,L93,CA98,ZC13} that in essence, decomposes a trajectory function into its constituent frequencies. Also, it is important to note that a typical galaxy has billions of stars, making simulations computationally expensive, therefore several analytical models are proposed based on Newtonian theory and General Relativity.

In Newtonian theory, globular and spherical galaxies are represented by the popular models of Plummer \citep{P11} and King \citep{K66}. Highly flattened axisymmetric galaxies are usually modeled by the so-called Toomre model \citep{T63}, while thickened-disk are modeled by Miyamoto and Nagai potentials \citep{MN75}. The Miyamoto-Nagai model is a stationary and axially symmetric potential, which despite the deviations in its density profile in comparison with realistic models, provides a simple and reasonable model for a disk. Concerning the galactic nuclei, most of the analytical models are spherical, finite mass, and its isotropic velocity dispersion can be calculated analytically \citep[see e.g.,][]{J83,Z85,H90,Z96}. Newtonian models for massive black hole nuclei have been also proposed aiming to reproduce the final stage for the evolution of active galactic nuclei \cite{R84} or to mimic realistically the nuclei of galaxies \citep{HR04}. The linearity of Newtonian equations allows the superposition of different models to study, e.g., galactic models composed of a disk and a bulge \citep{BT08}.

In General Relativity besides the well-known spherical solutions, several exact disk-like solutions have been proposed to model galaxies, including static thin disks \citep{BS68,MM69,UL04} and thick disks \citep{GL04}. In order to obtain these models is usual to start with the metric and calculate the energy-momentum from the metric. A different approach uses the energy-momentum tensor as a source of the Einstein equations but this is highly nontrivial. In the general relativistic framework, the composition of different models is not always possible due to the nonlinearity of the Einstein equations. Nevertheless, there are some special cases where the superposition of exact solutions is possible \citep{LL93,S02,S04}. A review of these and other relativistic models can be found in \citep{S02,VL05}.

There is compelling evidence of supermassive black holes in the center of several galaxies, including Milky Way \citep{LZH19}, where general relativity is found to provide a natural scenario for its study. On the other side, Newtonian models are computationally cheaper and can be easily built and analyzed. A balance between the two extremes can be achieved using Newtonian equations that include pseudo-potentials which reproduce some relativistic effects. In particular, some of the most renowned pseudo-Newtonian potentials have been introduced originally to study accretion disks in Black Holes \citep{PW80,SK99}. The simplest and at the same time practical of these potentials is the one derived by Paczy\'{n}ski and Wiita \citep{PW80}, which can correctly reproduce the locations of both the marginally bounded circular orbit and the last stable circular orbit of the Schwarzschild metric \citep{A09}.

Since is a well-known fact that pseudo-Newtonian potentials can have a remarkable effect on the dynamics of different systems \citep{SL06,DLG17,ZS19,ZDNT19}, in the present paper we shall use the pseudo-Newtonian Paczy\'{n}ski-Wiita potential \citep{PW80}, for modeling the central part of a galaxy hosting a supermassive black hole, while the flat thin disk is represented by the Miyamoto-Nagai potential. Aiming to obtain direct outcomes on how the mimicked relativistic properties of the supermassive black hole affect the character of the trajectories of the stars, we study the influence of the Schwarzschild radius and the angular momentum of a test particle on the orbit classification, i.e., collisional, ordered, and chaotic. Therefore, we would be able to obtain direct outcomes on how the relativistic properties of the supermassive black hole (such as the Schwarzschild radius) affect the character of the trajectories of the stars, by comparing results from previous works \citep[see e.g.,][]{Z14,ZC13}, where the central nucleus was modeled by using a simple Newtonian Plummer potential \citep{P11}. Regular and chaotic orbits will be identified by using the SALI method introduced by \citep{SABV04}, while regular orbits are further classified into families of regular orbits \citep{CA98,ZC13}.

The article is structured as follows: in Section \ref{mod} we present the main properties of the galaxy model. In the following Section \ref{cometh} we explain the computational methods we use for obtaining the classification of the orbits which is presented in Section \ref{clas}. Our article ends with Section \ref{conc}, where we emphasize the main outcomes of our analysis.

\section{Description of the galaxy model}
\label{mod}

Our galaxy model describes the motion of stars on the meridional $(R,z)$ plane and consists of two components. The first component is a flat thin disk, represented by the Miyamoto-Nagai potential \citep{MN75}
\begin{equation}
\Phi_{\rm d}(R,z) = - \frac{G M_{\rm d}}{\sqrt{R^2 + \left(k + \sqrt{z^2 + s^2}\right)^2}},
\label{disk}
\end{equation}
where, $M_{\rm d}$, $k$, and $s$ are the mass, the scale length and the scale height of the disk, respectively.

For the description of the central black hole, we use the Paczy\'{n}ski-Wiita potential \citep[see e.g.,][]{PW80,A09}.
\begin{equation}
\Phi_{\rm bh}(R,z) = - \frac{G M_{\rm bh}}{\sqrt{R^2 + z^2} - r_s},
\label{bh}
\end{equation}
where $M_{\rm bh}$ is the mass of the black hole. This is a very simple and practical pseudo-Newtonian potential, which however can realistically replicate and model general relativistic effects, associated with the motion of test particles around a non-spinning black hole. Moreover, the Paczy\'{n}ski-Wiita potential alters the classical Newtonian potential $1/r$ to $1/(r - r_s)$, thus introducing the relativistic Schwarzschild radius $r_s$.

Due to the fact that the total potential $\Phi_t(R,z) = \Phi_{\rm d} + \Phi_{\rm bh}$ has an axial symmetry, we have the conservation of the $z$-component $(L_z)$ of the total angular momentum. This automatically implies that the motion of the test particle can be described using the effective potential
\begin{equation}
\Phi_{\rm eff}(R,z) = \Phi_t(R,z) + \frac{L_z^2}{2R^2},
\label{veff}
\end{equation}
and the set of the following equations of motion
\begin{align}
&\ddot{R} = - \frac{\partial \Phi_{\rm eff}}{\partial R},\\
&\ddot{z} = - \frac{\partial \Phi_{\rm eff}}{\partial z}.
\label{eqmot}
\end{align}

The energy integral of the system is given through the Hamiltonian
\begin{equation}
H(R,z,\dot{R},\dot{z}) = \Phi_{\rm eff}(R,z) + \frac{1}{2} \left(\dot{R}^2 + \dot{z}^2 \right) = E,
\label{ham}
\end{equation}
where $E$ is the test particle's conserved value of the total orbital energy. Here it should be noted that the test particle can access those areas of the phase space where $E \geq \Phi_{\rm eff}$.

In our computations we use a system of units in which the unit of length is 1 kpc, the unit of time is $10^6$ yr (1 Myr), the unit of mass is $2.22508 \times 10^{11} {\rm M}_\odot$, the velocity unit is 978.564 km s$^{-1}$, the angular momentum unit (per unit mass) is 978.564 km kpc$^{-1}$ s$^{-1}$, while the unit of energy (per unit mass) is $9.576 \times 10^5$ km$^2$s$^{-2}$. Then for the gravitational constant we have that $G = 1$. In the above units the values of the constant parameters are: $M_{\rm d} = 0.5$ (corresponding to about $10^{11}$ M$_\odot)$, $k = 3$, and $s = 0.175$. On the other hand, the parameters $L_z$ and $M_{\rm bh}$ are treated as free parameters.

The Schwarzschild radius is related to the mass of the black hole through the relation
\begin{equation}
r_s = \frac{2 G M_{\rm bh}}{c^2}.
\label{rad}
\end{equation}

In the adopted system of galactic units, the value of the speed of light is roughly equal to 300 velocity units. Moreover, the values of the mass of the central black hole lie in the interval $[10^{8} {\rm M}_\odot, 10^{10} {\rm M}_\odot]$, which is a typical range of values in disk galaxies \citep{BT08}. Consequently, from Eq. (\ref{rad}) we derive that the values of the Schwarzschild radius lie in the interval $[10^{-8}, 10^{-6}]$. However, such low values will certainly induce computational malfunctions, since almost all numerical integrators cannot cope with motion extremely close to the singularity\footnote{Our previous experience indicates that the Bulirsch-Stoer integrator we use in our study can adequately follow the evolution of a test particle up to a radius equal to $10^{-5}$ around the singularity, while for smaller radii it displays several well known malfunctions, such as time step ``overflow".}. On this basis, in order to eliminate all ill behaviors during the numerical integration, we define a close encounter radius $R_c$ around the singularity as $R_c = 10^3 r_s$. At this point, it should be noted that although there are some studies in which the motion of test particles can reach a closer position to the galactic center (see e.g. \cite{SS13}), in the current study the close encounter radius does not affect the orbital dynamics of the test particle because the numerical artifact solely stops the numerical integration, therefore in practice, its direct effect is the loss of information in the basin diagrams around the origin in a disk of the order $[10^{-5}, 10^{-3}]$ units of length for $M_{bh}= [10^{8} {\rm M}_\odot, 10^{10} {\rm M}_\odot]$, respectively.

\begin{figure*}[!t]
\centering
\resizebox{\hsize}{!}{\includegraphics{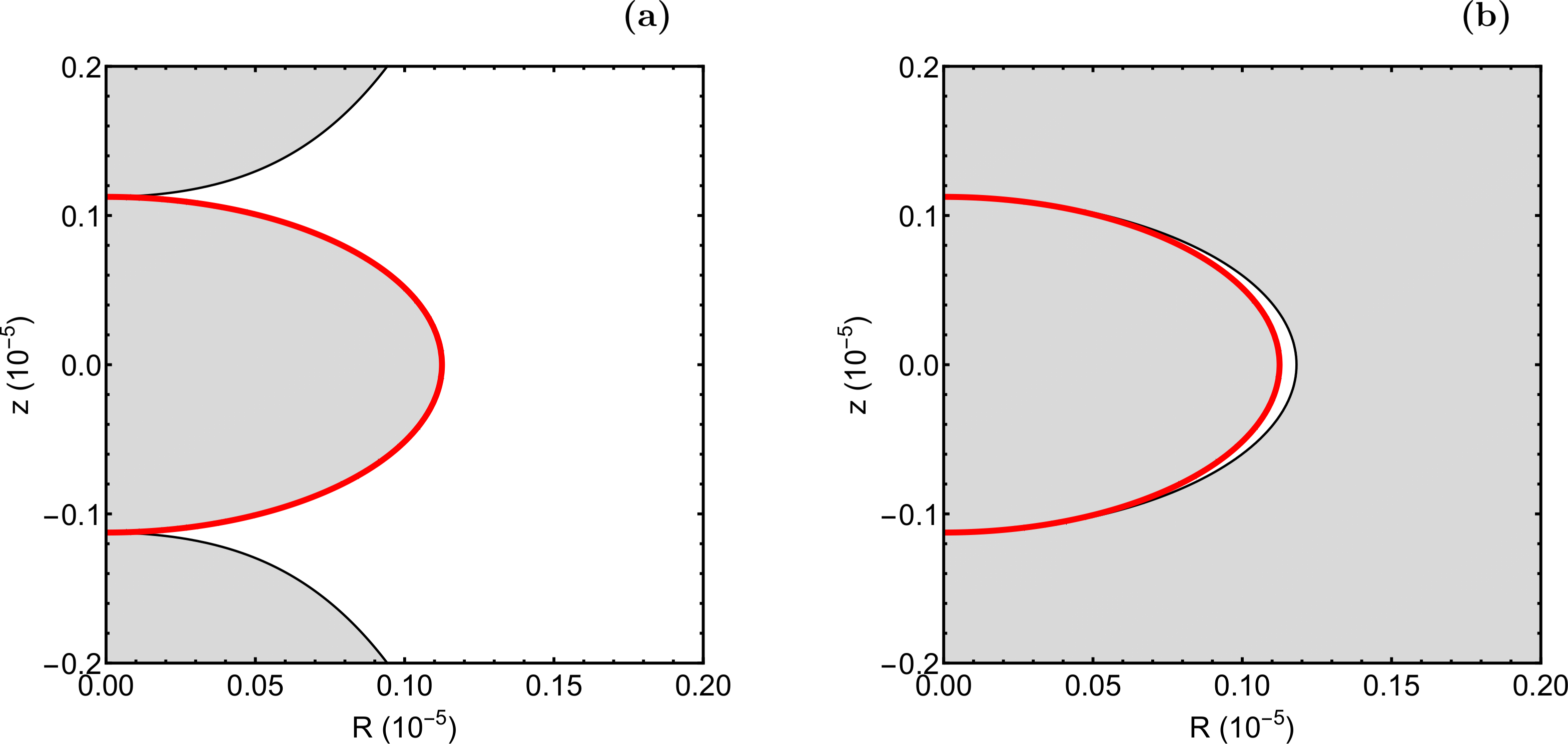}}
\caption{The geometry of the zero velocity curves (black lines), near the central region of the galaxy, for $M_{\rm bh} = 0.05$ and (a): $L_z = 0.1$, (b): $L_z = 0.5$. The energetically allowed and forbidden regions of motion are indicated using white and gray color, respectively, while the red line corresponds to the Schwarzschild radius.}
\label{conts}
\end{figure*}

\begin{figure}[!t]
\centering
\resizebox{\hsize}{!}{\includegraphics{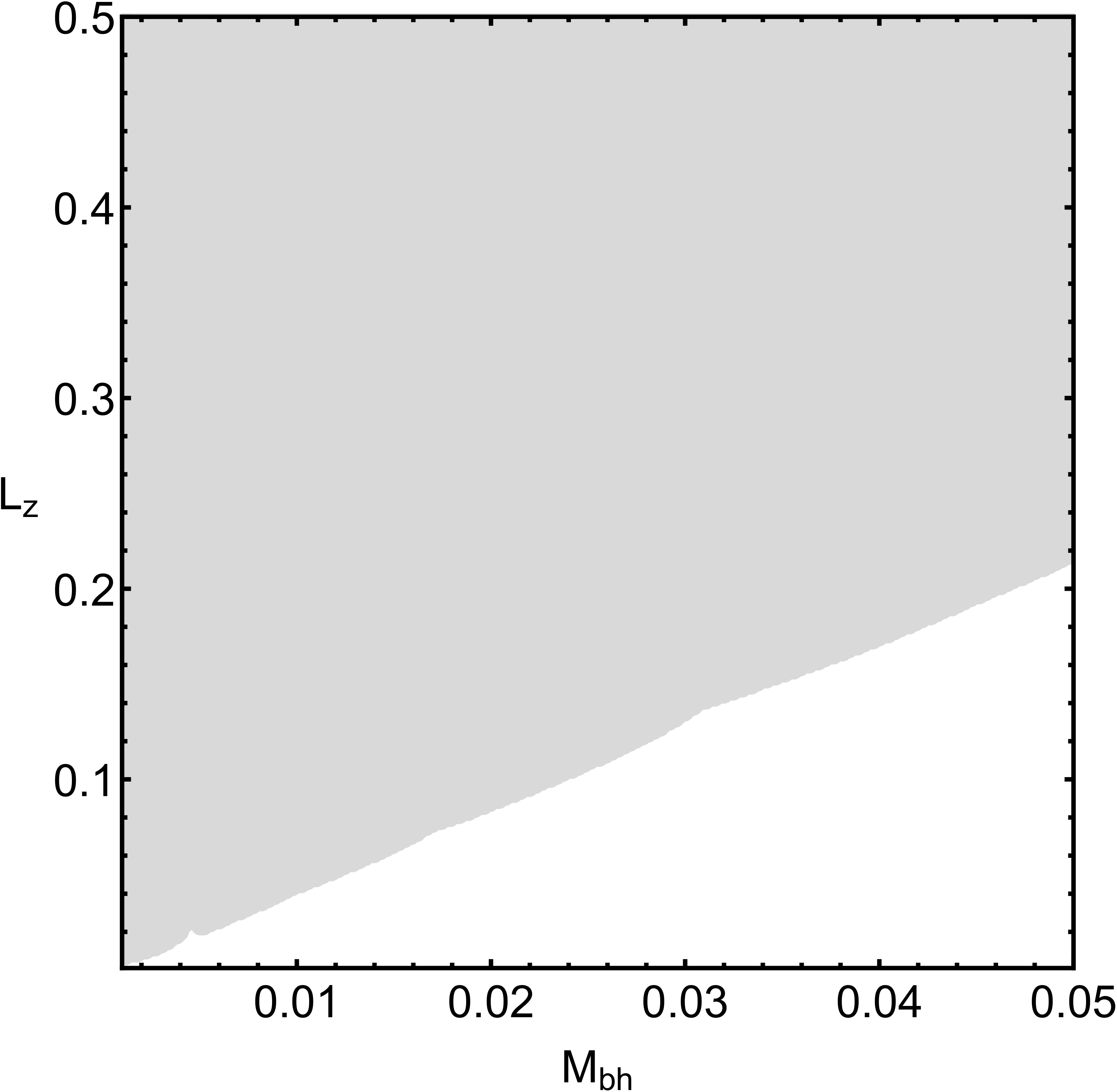}}
\caption{White areas correspond to open ZVCs, where collision to the central black hole is possible, while gray areas denote closed ZVCs, where the test particle is not energetically allowed to approach the Schwarzschild radius.}
\label{lims}
\end{figure}

In \citet{ZDG18} we have shown that as long as $r_s > 0$ there exists an additional energetically forbidden circular region around the black hole. Furthermore, the radius of the circular region is equal to the Schwarzschild radius. The interesting question is the following: is the test particle always able to reach the central black hole? The answer is no. Whether the test particle can approach the Schwarzschild radius or not, strongly depends on the value of its angular momentum. More precisely, if its angular momentum is low enough then it can approach the Schwarzschild radius of the black hole and display a collision event. On the other hand, if its angular momentum is relatively high then it can never reach the boundaries of the black hole. In panel (a) of Fig.~\ref{conts}, we present the first case, where the approach to the central black hole is possible. In this case, the zero-velocity curves (ZVCs), defined as $\Phi_{\rm eff}(R,z) = E$, are open, thus allowing the test particle to approach the black hole. In panel (b) of the same figure, we see the case where the test particle cannot reach the central region of the galaxy. This is because in this case the ZVCs are closed, thus acting as potential barriers. In Fig.~\ref{lims} we show the regions of open (white) and closed (gray) ZVCs, as a function of $(L_z, M_{\rm bh})$.

\section{Computational methodology}
\label{cometh}

\begin{figure*}[!t]
\centering
\resizebox{0.7\hsize}{!}{\includegraphics{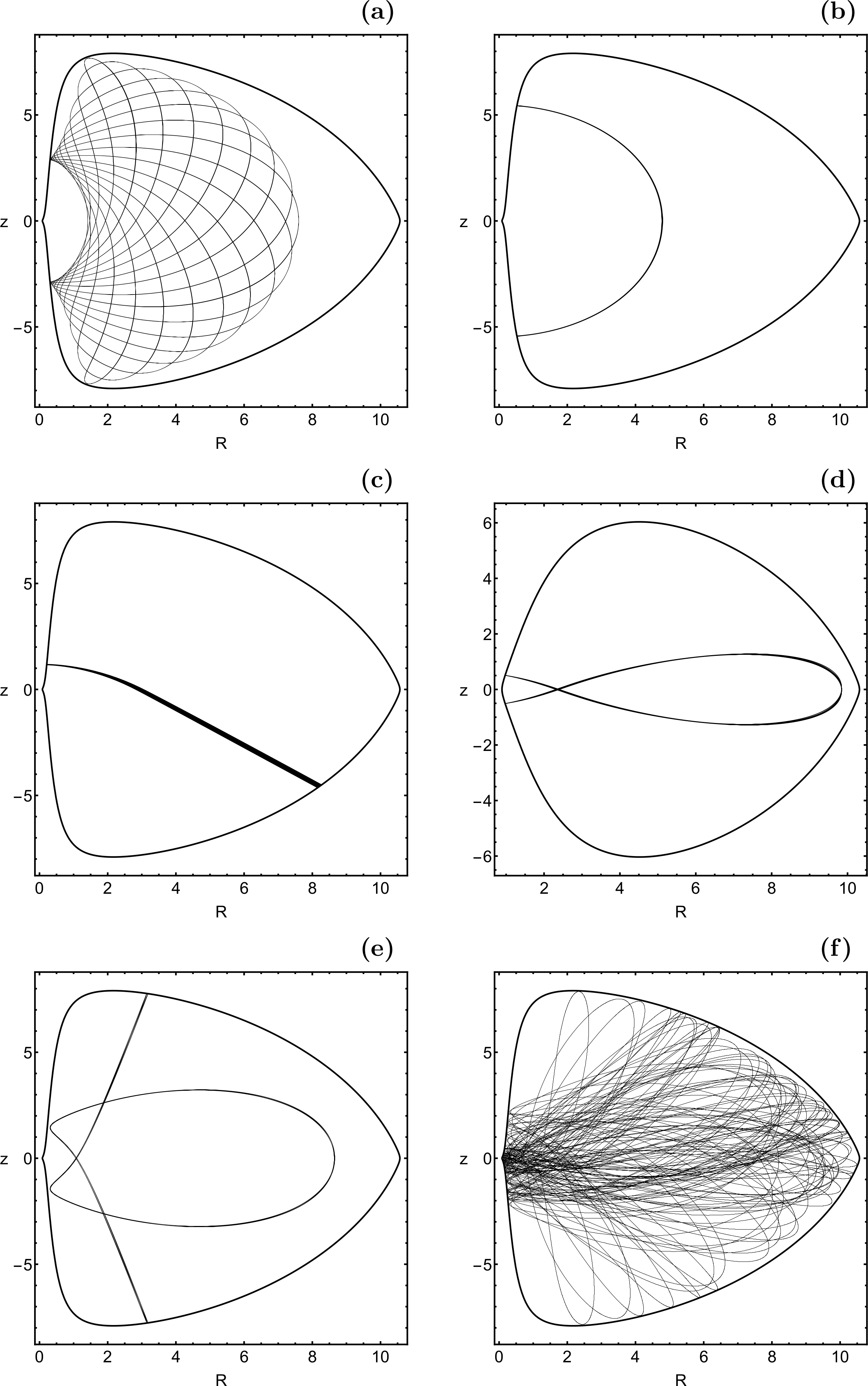}}
\caption{A collection showing the most important types of motion, when $M_{\rm bh} = 0.05$. (a): box orbit; (b): 2:1 orbit; (c): 1:1 orbit; (d): 2:3 orbit; (e): 4:3 orbit; (f): chaotic orbit.}
\label{orbs}
\end{figure*}

Our work aims to determine how the mass $M_{\rm bh}$ of the black hole, along with the value of the angular momentum $L_z$, affects the character of motion of the test particles. For this task, we perform an orbit classification, by numerically integrating sets of $1024 \times 1024$ initial conditions of orbits. In all cases, the value of the total orbital energy is fixed to $E = -0.05$, which corresponds to the maximum possible value of $R$ coordinate (about 10 kpc for a typical disk galaxy \citep[see e.g.,][]{BT08}). We decided to keep fixed the value of the energy because its influence on the character of orbits moving in the meridional plane has already been revealed in \citet{Z16}.

All initial conditions of the trajectories were numerically integrated for $10^4$ time units, which correspond to about 1 Hubble time (14 billion years) while using a constant time step, equal to $10^{-3}$. During the numerical integration, we also compute the Smaller Alignment Index (SALI) \citep{S01}, so as to be able to distinguish between chaotic and regular motion. This method allows classifying the orbits according to the numerical value obtained after evolving two orthonormal deviation vectors $\vec{w_{1}}$ and $\vec{w_{2}}$, which are periodically normalized to prevent overflow errors. More precisely, if SALI $>10^{-4}$ the orbit is categorized as regular, while if SALI $<10^{-8}$ it is classified as chaotic, or if the result belongs to the interval $10^{-4}<$ SALI $<10^{-8}$, it is classed as sticky, and the orbit requires a longer time of integration to be classified  \footnote{We refer the reader to \citep{S01} for a detailed explanation of the method.}. The index is formally defined as $\mathrm{SALI} \equiv \min \left(d_{-}, d_{+}\right)$, with
\begin{equation}
d_{\mp} \equiv \left\| \frac{\vec{w}_{1}}{\left\|\vec{w}_{1}\right\|} \mp \frac{\vec{w}_{2}}{\left\|\vec{w}_{2}\right\|} \right\|.
\end{equation}

Once an orbit was classified as a regular one, then we performed an additional categorization, thus classifying regular initial conditions into regular families. Our analysis indicates that several types of regular orbits appear in our disk galaxy model, while the most important ones are: (i) box orbits; (ii) 1:1 orbits; (iii) 2:1 orbits; (iv) 2:3 orbits; (v) 4:3 orbits; and (vi) higher resonant orbits. Specifically, in all studied cases, the relative percentage of higher resonant orbits is always less than 0.1\% and therefore we may argue that these orbits do not have any significant contribution to the overall orbital dynamics of the galaxy. The notation of orbits (box orbits and $n:m$ resonant orbits) in the axially symmetric galaxy model is according to the initial works of \citet{CA98} and \citet{ZC13}. In panels (a)-(e) of Fig.~\ref{orbs} we provide the shapes of the main regular orbits of the system, while in panel (f) of the same figure we give an example of a typical chaotic orbit when $M_{\rm bh} = 0.05$. In all cases, the ZVC is the black thick curve circumscribing the orbits. For all regular types of trajectories, we tried to present regular orbits with starting conditions very close to the respective parent periodic orbits, to be able to clearly recognize their shapes.

A double-precision Bulirsch-Stoer integrator, written in \verb!FORTRAN 77! \citep[e.g.,][]{PTVF92}, has been applied for the numerical integration of the starting conditions of the trajectories. During the numerical integration, the test particle's total energy was sufficiently conserved, while the corresponding observed error was of the order of $10^{-14}$. For the classification of the initial conditions (per grid) the required CPU time, using a Quad-Core Intel i7 4.0 GHz processor was varying between 5 and 22 hours, strongly depending on the number of collision orbits. Version 12.1 of Mathematica$^{\circledR}$ software has been deployed for developing all the graphical illustration of the article.

\section{Orbit classification}
\label{clas}

\begin{figure*}[!t]
\centering
\resizebox{\hsize}{!}{\includegraphics{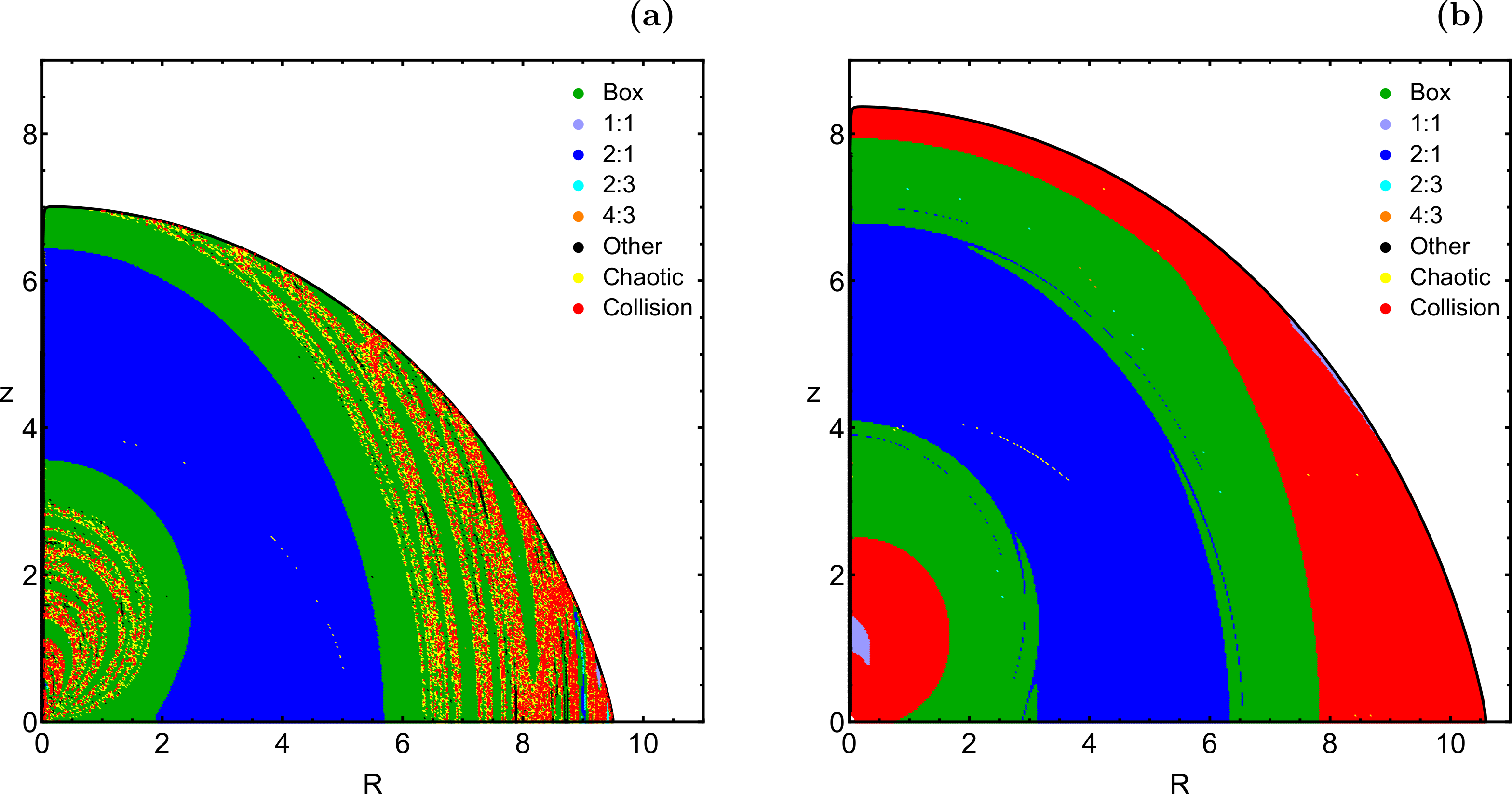}}
\caption{Basin diagrams on the $(R,z)$ plane for $L_z = 0.001$, with (a): $M_{\rm bh} = 0.0005$ and (b): $M_{\rm bh} = 0.05$.}
\label{rz1}
\end{figure*}

\begin{figure*}[!t]
\centering
\resizebox{\hsize}{!}{\includegraphics{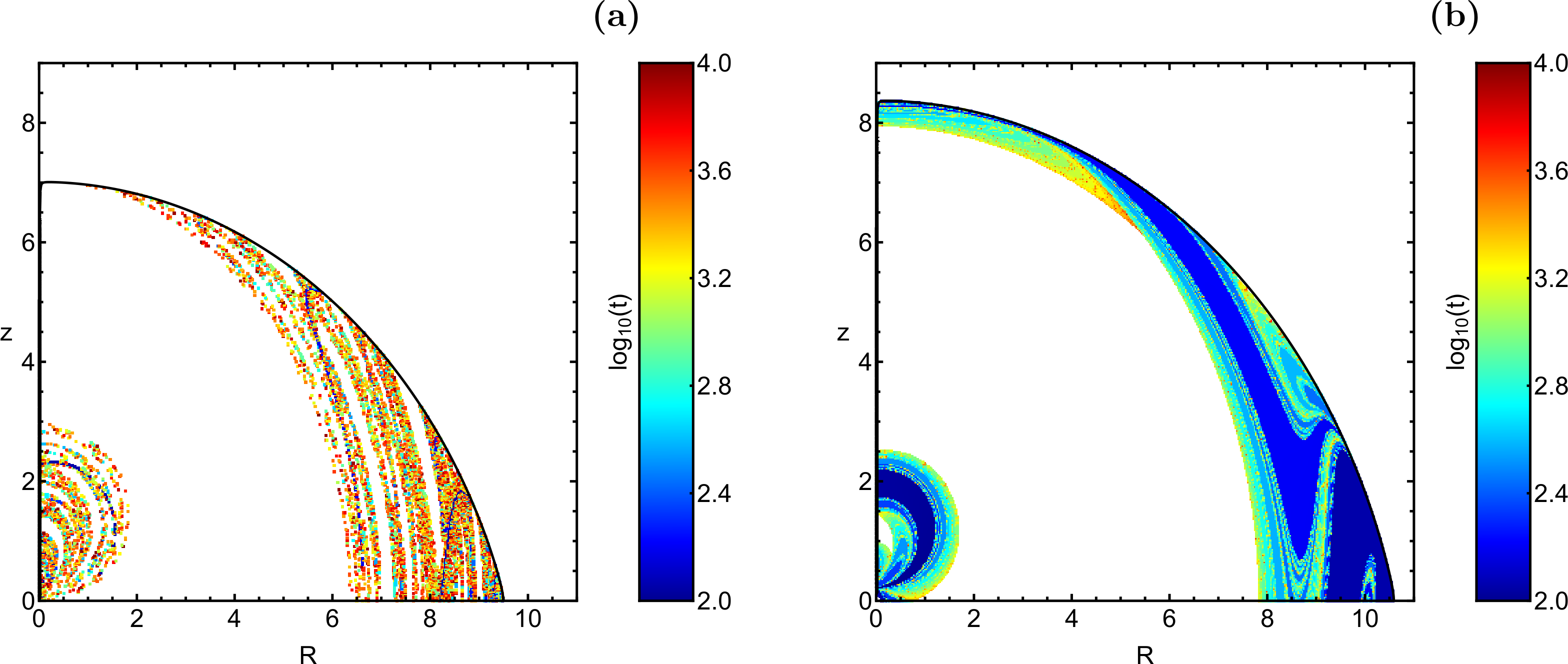}}
\caption{Distribution of the collision times for $L_z = 0.001$, with (a): $M_{\rm bh} = 0.0005$ and (b): $M_{\rm bh} = 0.05$.}
\label{tcol}
\end{figure*}

\begin{figure*}[!t]
\centering
\resizebox{\hsize}{!}{\includegraphics{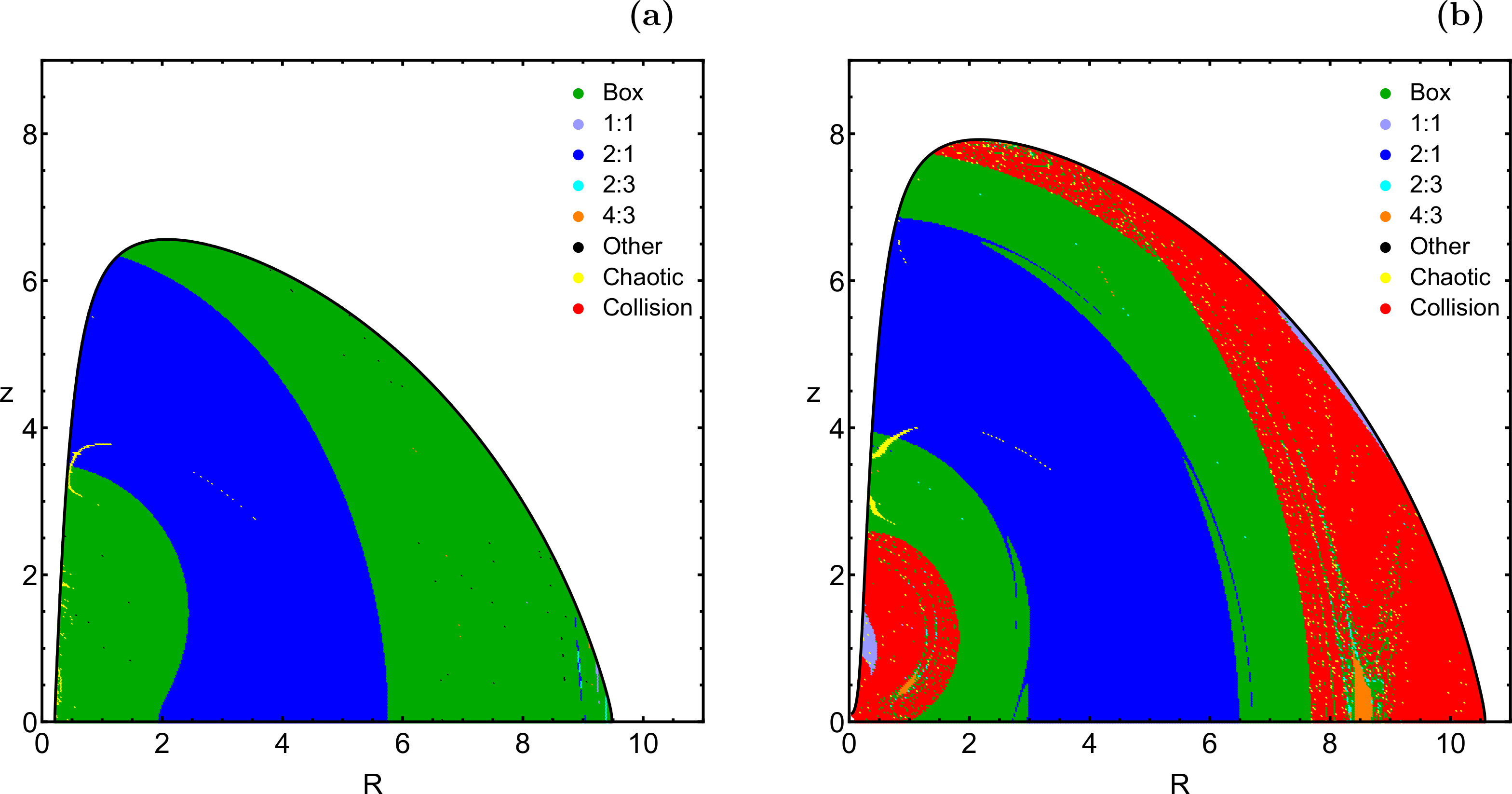}}
\caption{Basin diagrams on the $(R,z)$ plane for $L_z = 0.1$, with (a): $M_{\rm bh} = 0.0005$ and (b): $M_{\rm bh} = 0.05$.}
\label{rz2}
\end{figure*}

\begin{figure*}[!t]
\centering
\resizebox{\hsize}{!}{\includegraphics{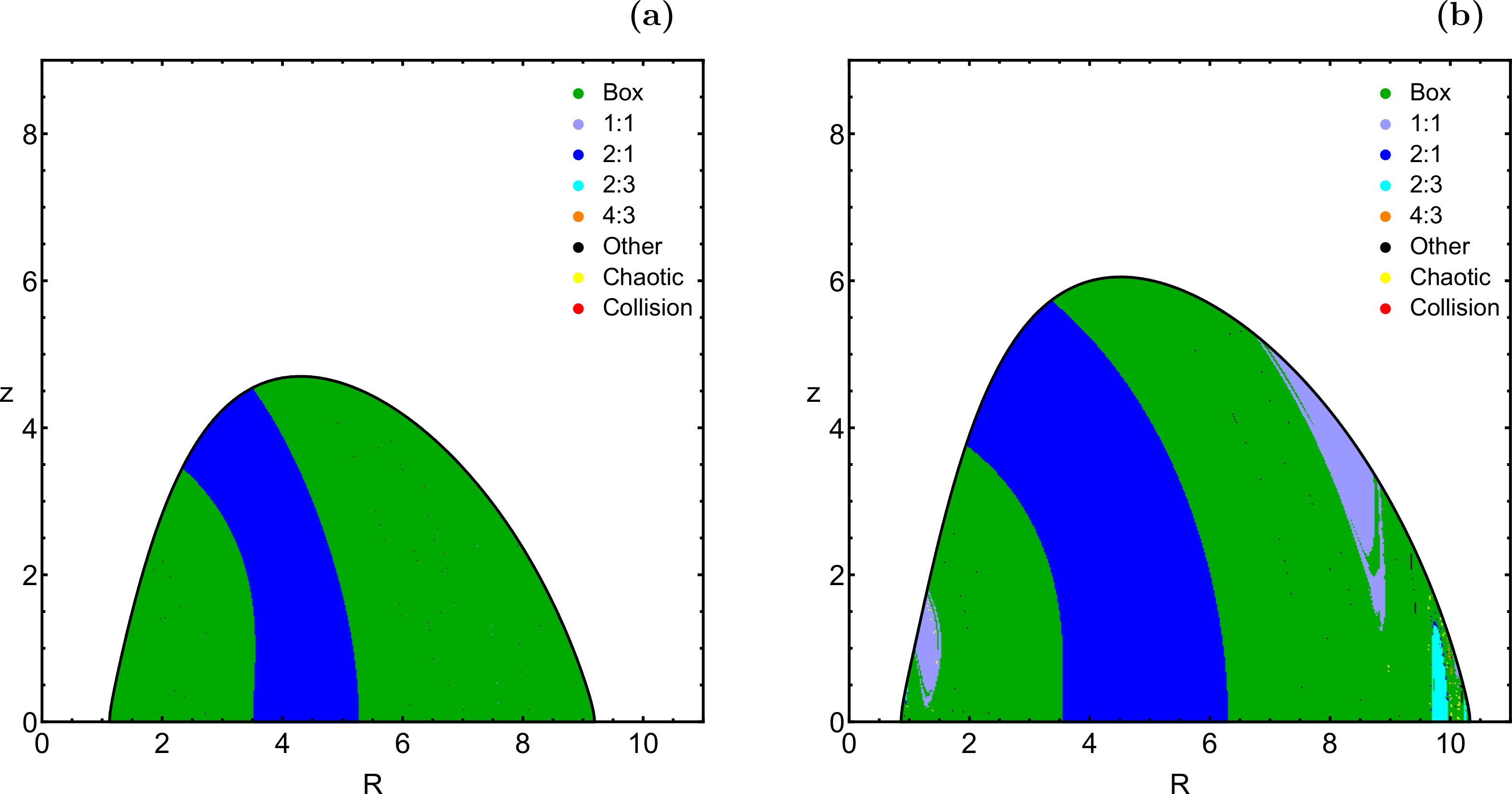}}
\caption{Basin diagrams on the $(R,z)$ plane for $L_z = 0.5$, with (a): $M_{\rm bh} = 0.0005$ and (b): $M_{\rm bh} = 0.05$.}
\label{rz3}
\end{figure*}

\begin{figure*}[!t]
\centering
\resizebox{\hsize}{!}{\includegraphics{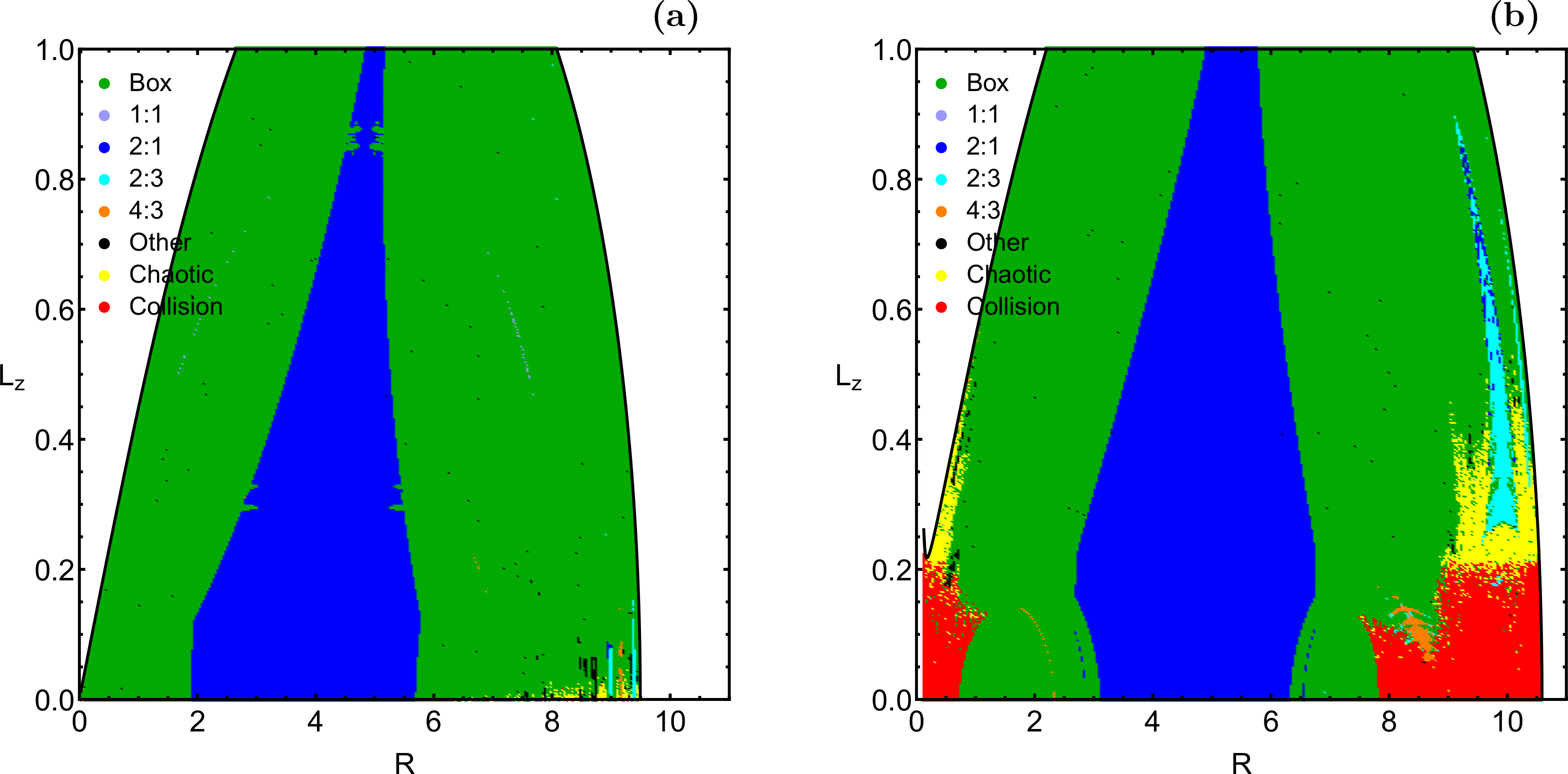}}
\caption{Basin diagrams on the $(R,L_z)$ plane, with (a): $M_{\rm bh} = 0.0005$ and (b): $M_{\rm bh} = 0.05$.}
\label{rlz}
\end{figure*}

\begin{figure*}[!t]
\centering
\resizebox{\hsize}{!}{\includegraphics{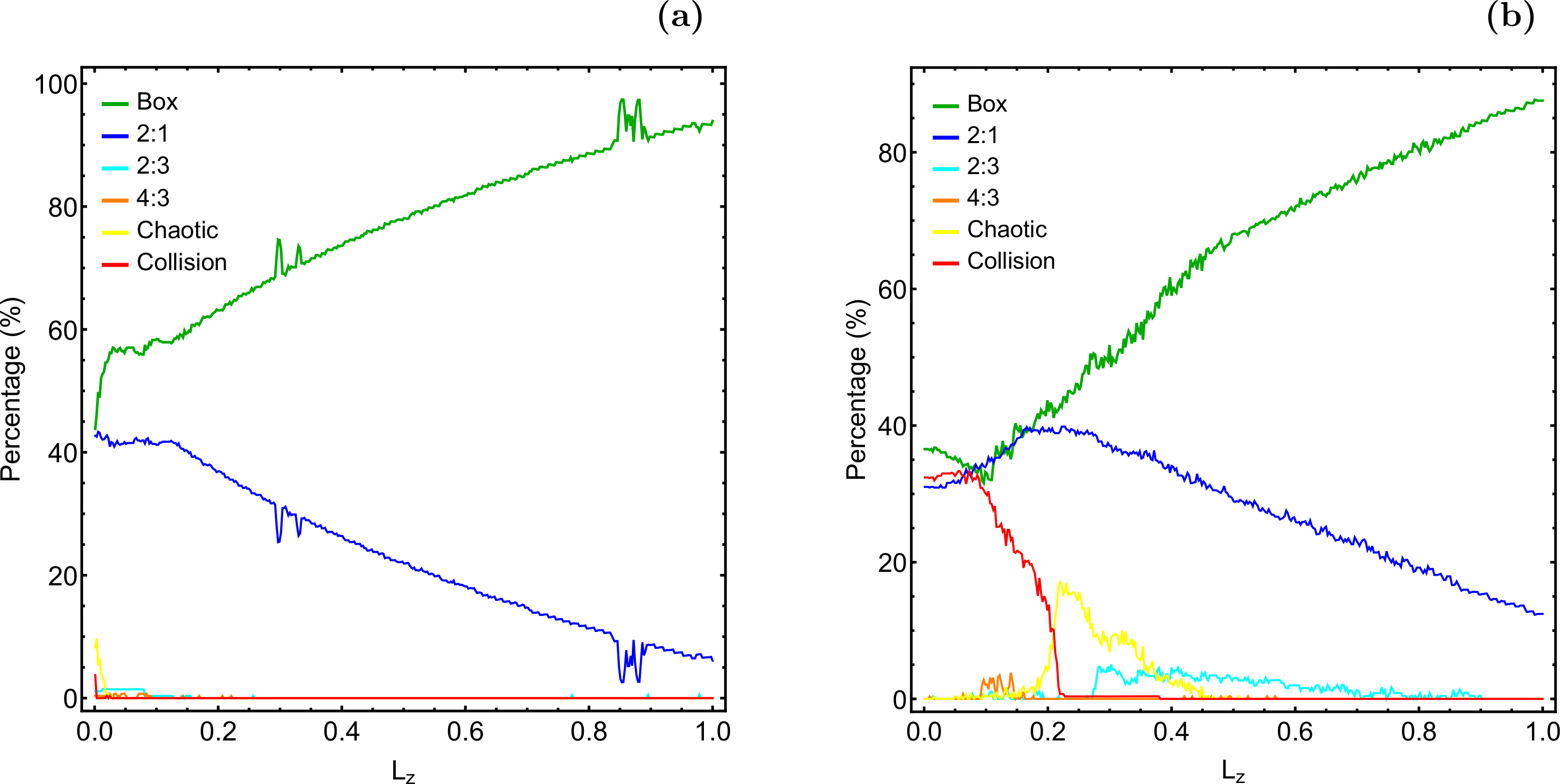}}
\caption{Evolution of the percentages of all main types of orbits on the $(R,L_z)$ plane, as a function of the angular momentum, when (a): $M_{\rm bh} = 0.0005$ and (b): $M_{\rm bh} = 0.05$.}
\label{percs}
\end{figure*}

In this section, we will present the nature of the motion of the test particle as a function of the mass of the black hole and the angular momentum. For illustrating the orbital properties of the galactic system, hereafter we shall follow the graphical approach introduced in \citet{N04,N05}, thus presenting color-coded basin diagrams, where each pixel corresponds to a unique trajectory and it is colored according to the final state of the test particle.

All trajectories have initial conditions $(R_0,z_0)$, while for a given total orbital energy $E_0$ the initial velocities are given by
\begin{align}
&\dot{R_0} =   \frac{z_0}{d_0}f(R_0,z_0;E_0),\\
&\dot{z_0} = - \frac{R_0}{d_0}f(R_0,z_0;E_0),
\label{vel}
\end{align}
where
\begin{equation}
f(R_0,z_0;E_0) = \sqrt{2\left(E_0 - \Phi_{\rm eff}(R_0,z_0) \right)},
\end{equation}
while $d_0 = \sqrt{R_0^2 + z_0^2}$.

\subsection{Low angular momentum}
\label{ss1}

We begin our analysis, considering the case where the test particle (star) takes low values of the angular momentum. In Fig.~\ref{rz1}(a-b) we present the basin diagrams on the $(R,z)$ plane, when $L_z = 0.001$. We consider only the $z > 0$ part because due to the axial symmetry of the galaxy the orbital structure of the $z < 0$ part is mirror-symmetrical. Panel (a) corresponds to the case where the mass of the black hole is $M_{\rm bh} = 0.0005$. The first type of motion which we encounter as we move away from the center is the box orbits. In fact, there are two main basins occupied by initial conditions of box orbits, while between these basins we have the presence of the 2:1 resonant orbits. Both basins of box orbits are ``polluted" by a mixture of chaotic and collision orbits, while tiny stability islands of higher resonant orbits are also present. Very close to the boundaries of the ZVC we observe the presence of 2:1 and 1:1 resonant orbits, while there are no signs of other resonant orbits, such as 4:3.

The case with $M_{\rm bh} = 0.05$ is displayed in panel (b) of Fig.~\ref{rz1}. Here we see that the pollution of collision orbits we have seen in panel (a) is much stronger leading to well-defined basins of collision. Inside the collision basin near the central black hole, we observe the presence of a stability island corresponding to 1:1 resonant orbits. On the other hand, higher resonant orbits seem to be completely absent for this value of $M_{\rm bh}$. Moreover, all initial conditions belonging to chaotic orbits (according to the value of SALI) lead, sooner or later, to a collision with the central black hole, and therefore there are no remaining trapped chaotic trajectories.

In order to illustrate more clearly the difference between the two cases regarding the collision orbits, we present in Fig.~\ref{tcol}(a-b) the distributions of the corresponding collision times of the trajectories. In panel (a), corresponding to $M_{\rm bh} = 0.0005$ one can see that the vast majority of the starting conditions need a significant amount of time (more than 2500 time units) for reaching the central black hole. However, we distinguish some very thin basins of initial conditions for which the corresponding collision time is very low (less than 300 time units). On the other hand, in panel (b) of Fig.~\ref{tcol} we see that almost all the initial conditions lead to a collision with the black hole within the first 1000 time units of the numerical integration. Therefore, one may reasonably ask: why the orbits collide so quickly in the second case? Fortunately, the answer is very easy. In both cases, the ZVCs are open near the center and therefore the test particle can approach the central black hole. The main difference is that in the second case (where $M_{\rm bh} = 0.05$) the Schwarzschild radius is 100 times larger which implies that the collision channel is much wider. Consequently, the test particles need significantly less time to find the open throat and collide with the singularity.

\subsection{Intermediate angular momentum}
\label{ss2}

The second case under investigation concerns test particles moving with intermediate values of angular momentum. The basin diagrams on panels (a) and (b) of Fig.~\ref{rz2} show the orbital structure of the meridional $(R,z)$ plane, when $L_z = 0.1$. Panel (a) corresponds to $M_{\rm bh} = 0.0005$. Here we see that collision motion is not possible. This is also seen in the diagram of Fig.~\ref{lims}, where one can extract the valuable information that for $(M_{\rm bh},L_z) = (0.0005,0.1)$ the corresponding ZVCs are closed. We may argue that for this set of values of $M_{\rm bh}$ and $L_z$ almost the entire $(R,z)$ plane is covered by box and 2:1 orbits, while all other types of orbits are extremely limited or even absent.

For a much higher value of the mass of the black hole, we see in panel (b) of Fig.~\ref{rz2} that extended collision basins emerge close and far from the central singularity, thus leading to a considerable reduction of the area corresponding to box orbits. Besides, one can see that now the test particle can perform also 1:1 and 4:3 resonant orbits.

\subsection{High angular momentum}
\label{ss3}

The last scenario under consideration is the case with high angular momentum stars. Fig.~\ref{rz3}(a-b) shows the basins diagrams on the $(R,z)$ plane, when $L_z = 0.5$. Now we observe that the value of the mass of the black hole affects only resonant orbits. This implies that high angular momentum stars, moving on box and 2:1 orbits are hardly affected by the central black hole. Furthermore, this type of stars cannot even approach the singularity, taking into account that for such high values of the angular momentum the ZVCs are always closed, thus not allowing collision with the black hole. Another interesting fact is that the 1:1 and 2:3 resonant types of motion seem to be more prominent, compared to all other previous cases, while on the other hand, the 4:3 resonant orbits are barely noticeable. Moreover, it should be noted that with increasing value of the angular momentum, while keeping constant the value of the total orbital energy, the energetically allowed region on the meridional $(R,z)$ plane is considerably reduced.

\subsection{Overview analysis}
\label{over}

It would be very informative to present from another perspective the influence of the angular momentum $L_z$ as well as of the mass of the black hole $M_{\rm bh}$ on the nature of the motion of the test particles.

In Fig.~\ref{rlz}(a-b) we present the character of the trajectories, with initial conditions on the $(R,L_z)$ plane, while for all orbits we set $z_0 = 0$. It is seen, that when the value of the mass of the black hole is low, collision orbits are very limited and appear only for very low values of the angular momentum $(L_z < 0.01)$, that is when stars can approach very close to the galactic center. On the other hand, when $M_{\rm bh}$ is high enough collision is still possible, even for relatively large angular momentum values $(L_z < 0.1)$. Box orbits occupy most of the $(R,L_z)$ plane, while the second most populated type of motion is the 2:1 resonant family, which lies around 4-5 kpc. Furthermore, one can see, that for $M_{\rm bh} = 0.05$, the 4:3 resonant orbits appear at about $L_z = 0.1$, while the 2:3 resonant orbits exist mainly for $L_z > 0.2$. Isolated initial conditions of higher resonant orbits can be observed scattered inside the area of box orbit. 

The parametric evolution of the percentages of all the main types of orbits, as a function of the angular momentum, is given in panels (a) and (b) of Fig.~\ref{percs}. From this figure, it is seen that the orbital content of the system is affected by the value of the mass of the black hole mainly for low values of the angular momentum $(L_z < 0.2)$, while for higher values the pattern is almost unaffected by the shift on the value of $M_{\rm bh}$. We also see, that for large angular momentum values $(L_z \to 1)$ box orbits completely dominate, while at the same time resonant orbits, such as the 2:3 and 4:3 orbital families disappear.

\begin{figure}[!t]
\centering
\resizebox{\hsize}{!}{\includegraphics{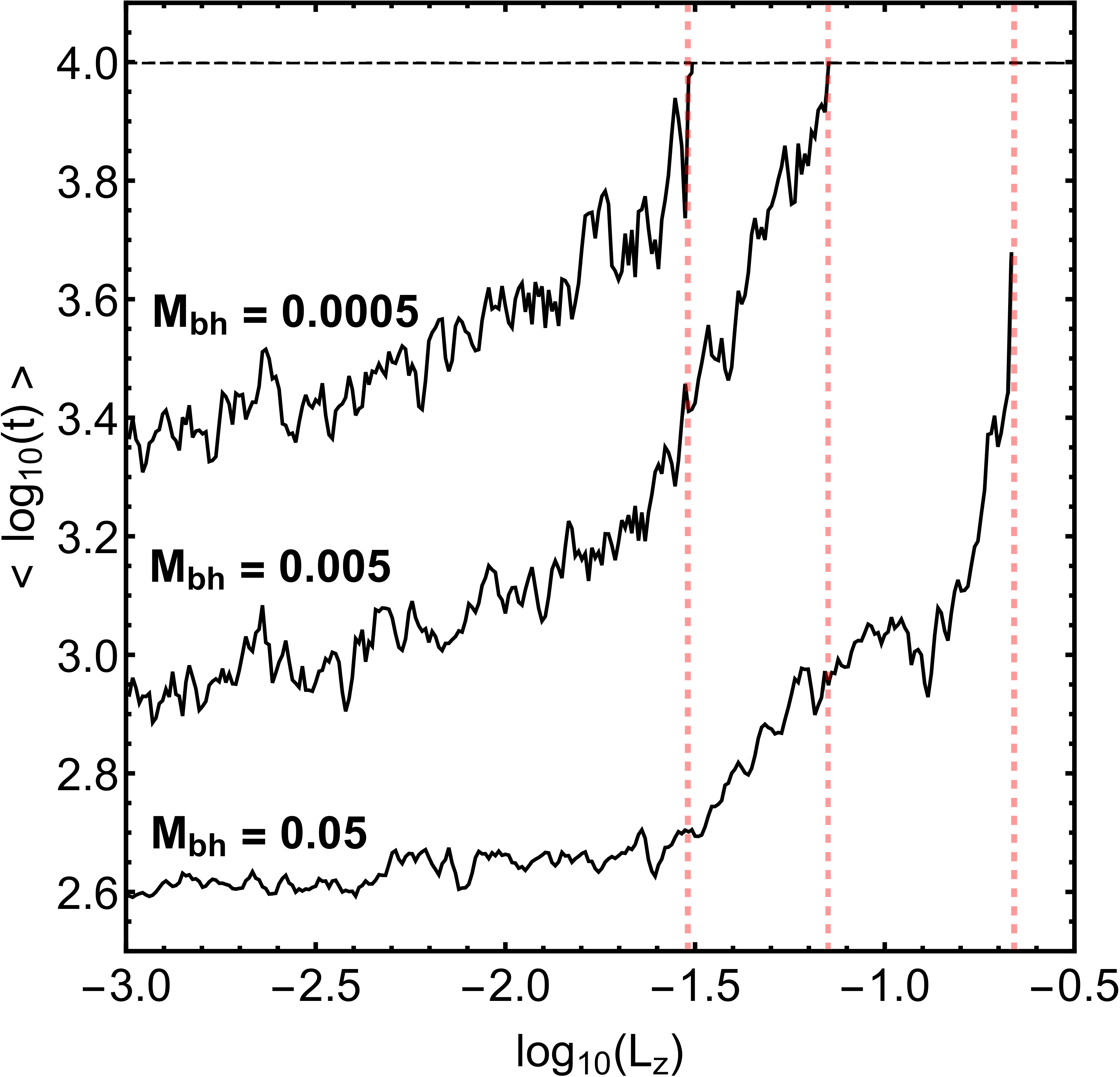}}
\caption{Evolution of the average collision time of the trajectories, with initial conditions on the $(R,L_z)$ plane, as a function of the angular momentum. The horizontal black dashed line indicates the total time of the numerical integration, while the vertical red dashed lines denote the maximum value of $L_z$, for which the ZVCs are open.}
\label{times}
\end{figure}

Before closing, we would like to present some additional information regarding the collision time of the trajectories and the connection between the angular momentum and the mass of the black hole. In Fig.~\ref{times} we display the evolution of the average collision of orbits as a function of the angular momentum, for three values of $M_{\rm bh}$. We see, that the average collision time is reduced, with increasing value of the mass of the black hole. Moreover, in all three cases, the collision time increases, with increasing value of the angular momentum, and tends to the total time of the numerical integration ($10^4$ time units). This is anticipated because with an increasing value of $L_z$ the width of the collision channel of the ZVCs, near the galactic center, is reduced. The vertical red dashed lines indicate the maximum values of the angular momentum for which a collision is possible (evidently, for higher values of $L_z$ the corresponding ZVCs are closed). We would also like to note that these critical values of the angular momentum, obtained through the numerical integration, coincide with the respective values obtained earlier (see Fig.~\ref{lims}), using analytical methods.

\section{Concluding remarks}
\label{conc}

In this work, we numerically investigated the motion of stars on the meridional $(R,z)$ plane of an axially symmetric disk galaxy model, containing a central supermassive black hole. For modeling the black hole we used the Paczy\'{n}ski-Wiita potential which can replicate important relativistic properties, such as the existence of the Schwarzschild radius $r_s$. After classifying extensive samples of initial conditions of trajectories, we managed to distinguish between collisional, ordered, and chaotic motion. Furthermore, all starting conditions of regular orbits are further classified into families of regular orbits. Our results are presented through modern color-coded basin diagrams on several types of two-dimensional planes, while we also monitored the evolution of the respective percentages.

The following list summarizes the main outcomes, regarding the motion of stars:
\begin{enumerate}
  \item We have seen that the parameters of both the angular momentum and the mass of the black hole determine if the stars can approach or not to the central singularity. In general terms, open ZVCs (which allow collision with the black hole) are present only when $L_z < 0.2$.
  \item For low values of the angular momentum, only the most basic types of orbits exist (such as box, 1:1, and 2:1), while higher resonant orbits (such as the 2:3 and the 4:3 orbital families) appear only for higher values of the $L_z$.
  \item For relatively low values of the mass of the black hole $(M_{\rm bh} \simeq 0.0005)$ we encountered the phenomenon of trapped chaos, where chaotic orbits remained trapped inside the ZVCs for at least one Hubble time, even though collision to the central black hole is energetically allowed. On the other hand, for high values of $M_{\rm bh}$, there is no evidence of trapped chaos and all chaotic orbits lead fast to a collision.
  \item Low angular momentum stars (with $L_z < 0.1$) are highly affected by the mass of the black hole. More specifically, for low values of $M_{\rm bh}$ they can either move on a box, 2:1, or chaotic orbit, while a collision is less possible. On the contrary, for high values of $M_{\rm bh}$ the possibility of collision is much higher, while regular motion seems to be shared between box and 2:1 types of orbits.
  \item High angular momentum stars (with $L_z > 0.5$) seem to be less affected by the value of the mass of the black hole since they cannot even approach the central regions of the galaxy. Our analysis indicates that the vast majority of high angular momentum stars move either on box or 2:1 trajectories.
\end{enumerate}

In two earlier works \citep{Z14,ZC13} we have also investigated the nature of motion in a disk galaxy model, using the same Miyamoto-Nagai model. However, the central nucleus was modeled, using a typical Plummer potential \citep{P11}, while in the present study we use a pseudo-Newtonian Paczy\'{n}ski-Wiita potential \citep{PW80}. Therefore, we can directly compare the differences between the classical Newtonian and the pseudo-Newtonian models. In both cases, the main types of regular orbits are the same, while the influence of the angular momentum seems to be similar. In particular, we see that high angular momentum stars move mainly on regular orbits, while low angular momentum stars can display chaotic motion, by passing close to the central region. The main difference between the two cases is the fact that in the case of the Plummer potential we encountered additional types of resonant orbits (e.g., 5:4 and 6:5). Thus, we may argue that in the pseudo-Newtonian case of the Paczy\'{n}ski-Wiita potential the orbital content of regular motion is not as rich as in the case of the Plummer potential. Additionally, in the case of the Plummer potential several types of resonant orbits (such as the 2:3, 4:3, 5:4 and 6:5) were simultaneously present. On the contrary, in the case of the Paczy\'{n}ski-Wiita potential, we observe a hierarchy regarding the appearance of resonant orbits. Specifically, for low values of $L_z$ we have the presence of the 4:3 orbits, while the 2:3 orbits appear for much higher values of the angular momentum. Finally, for the Newtonian case of the Plummer potential we encountered many more types of higher resonant orbits $n:m$ (with $n, m > 5$) inside the area of the box orbits, while in the case of the Paczy\'{n}ski-Wiita potential all these higher resonant orbits are completely absent.

Consequently, we believe that our findings can be relevant not only from a theoretical point of view but also from an astrophysical perspective. In particular, the identification of resonant orbits could shed lights on the formation and origin of kinematic moving groups (MGs) \citep{MPMPV16}, while the presence of chaotic or collisional orbits could certainly modify the structure and hence the evolution of the galaxy with the increase of the central black hole mass. Taking into account the positive and encouraging results of the present analysis, it is in our plans to expand our orbit classification into three dimensions. We would aim to determine how the mass of the black hole in combination with the angular momentum of the stars, affects their trajectories inside the three-dimensional $(x,y,z)$ space. Moreover, it would also be interesting to determine the influence of the mass of the black hole on the network of periodic orbits of the galaxy.

\section*{Acknowledgments}

This project was funded by the Deanship of Scientific Research (DSR) at King Abdulaziz University, Jeddah, Saudi Arabia, under grant no. KEP-17-130-38. The authors, therefore, acknowledge with thanks DSR technical and financial support. FLD acknowledges financial support from the Universidad de los Llanos. Our warmest thanks go to the anonymous referee for the careful reading of the manuscript as well as for all the apt suggestions and comments, which allowed us to improve both the quality and the clarity of the paper.

\end{document}